\documentclass[prl,aps,tightenlines,showpacs,twocolumn,floatfix,
unsortedaddress,superscriptaddress]{revtex4-1}

\usepackage{graphicx}
\usepackage{amsmath} 
\usepackage{color}

\begin{document}

\title{Ground-state Properties of Unitary Bosons: From Clusters to Matter}
\author{J. Carlson}
\email{carlson@lanl.gov}
\author{S. Gandolfi}
\email{stefano@lanl.gov}
\affiliation{Theoretical Division, Los Alamos National Laboratory,
Los Alamos, New Mexico 87545, USA}
\author{U. van Kolck}
\email{vankolck@ipno.in2p3.fr}
\affiliation{Institut de Physique Nucl\'eaire, CNRS/IN2P3,
Universit\'e~Paris-Sud, Universit\'e Paris-Saclay, 91406 Orsay, France}
\affiliation{Department of Physics, University of Arizona, Tucson, 
Arizona 85721, USA}
\author{S.~A. Vitiello}
\email{vitiello@ifi.unicamp.br}
\affiliation{Instituto de F\'isica Gleb Wataghin,
Universidade Estadual de Campinas, 13083-859 Campinas SP, Brazil}

\begin{abstract}

The properties of cold Bose gases at unitarity have been
extensively investigated in the last few years both theoretically
and experimentally. In this paper we use a family of interactions
tuned to two-body unitarity and very weak three-body binding
to demonstrate the universal properties of both clusters and matter.
We determine the universal properties of finite clusters up to 60 particles
and, for the first time, explicitly demonstrate the saturation of 
energy and density with particle number and compare with bulk properties.
At saturation in the bulk we determine
the energy, density, two- and three-body contacts and the 
condensate fraction.
We find that uniform matter is more bound than three-body
clusters by nearly two orders of magnitude, the two-body contact
is very large in absolute terms, 
and yet the condensate fraction is also very large, greater than 90\%.
Equilibrium properties of these systems may be experimentally accessible through
rapid quenching of weakly-interacting boson superfluids. 
\end{abstract}

\date{\today}

\pacs{}

\maketitle

\emph{Introduction:}
Strongly-interacting fermionic cold atoms
have been the subject of a great deal of study both theoretically and
experimentally across the BEC to BCS transition, and especially at
unitarity, where the two-body system has nearly a zero-energy bound
state~\cite{Giorgini:2008}. These systems are universal in that all properties,
including ground-state energy, superfluid pairing gaps, superfluid transition
temperatures, {\it etc.}, are obtained as a set of universal dimensionless
parameters multiplied by the Fermi energy or momentum of a free Fermi
gas at the same density.  Studies of bosonic superfluids, however,
have concentrated on the weakly-interacting regime described by the
Gross-Pitaevski mean-field equation. These systems are comparatively
simple
to study as they were the first to be cooled to very low temperatures
and their properties can be described in a mean-field picture.

It has been known for some time that short-range two- and three-body
interactions can be used to describe the low-energy properties of
small clusters of bosons.
To obtain universal properties, the
two-body interaction can similarly be taken to generate
a zero-energy dimer,
but a three-body interaction is required \cite{Bedaque:1998kg,Bedaque:1998km}
to avoid the so-called ``Thomas collapse'' \cite{Thomas:1935zz}
of three or more particles.  
The resulting discrete scale invariance leads to geometric
towers of states in systems with three \cite{Efimov:1970zz} and more 
\cite{Hammer:2006ct,Stecher:2009,Deltuva:2010xd,Gattobigio:2011ey,vonStecher:2011zz} 
bosons.
Many atomic and nuclear few-body systems fall into this universality class
\cite{Braaten:2004rn}. 

In this paper we demonstrate
that large clusters and bulk matter are stable with such interactions,
and similarly to the fermionic case described by a fairly simple set of
universal parameters. We provide the first estimates for the universal
parameters describing the ground-state energy, the equilibrium density,
two- and three-body contacts, and the condensate fraction of such
a system. 
Our calculations are the analog of those carried out for fermions in 
Refs. \cite{Carlson:2003zz,Chang:2004zza},
but here the universal parameters are directly related
to the properties of the three-body system, {\it i.e.} its energy and 
radius.  These bosonic universal
properties may be accessible through cold-atom experiments, including
those studying rapid quenching from weakly-interacting Bose condensates.

\emph{Interaction and Method:}
The Hamiltonian we consider is
\begin{equation}
H \ = \ -\frac{\hbar^2}{2m} \sum_i \nabla_i^2 
+ \sum_{i<j} V_{ij} + \sum_{i<j<k} V_{ijk},
\end{equation}
where the first term is the non-relativistic kinetic energy, the second
the attractive short-range interaction tuned to infinite scattering
length, and the last term is a repulsive three-body contact interaction
tuned to produce a weakly-bound trimer.  For zero-range interactions
universality has been demonstrated in Ref. \cite{vanKolck:1998bw}. 
For this study we employ
finite-range two- and three-body interactions, keeping the range of these
interactions much smaller than the size of the weakly-bound trimer.
For unitarity bosons this restriction is very stringent, as we shall
see.  The interaction must also be much shorter ranged than the average
interparticle spacing in the bulk, which is an order of magnitude smaller
than the three-body cluster size.

Here we employ Gaussian two- and three-body interactions:
\begin{align}
&V_{ij} \ = \ V_2^0\frac{\hbar^2}{m}\mu_2^2 \, 
\exp [ - (\mu_2 r_{ij})^2/2 ] \,, \\
&V_{ijk} \ = \ V_3^0\frac{\hbar^2}{m}
\left(\frac{\mu_3}{2}\right)^2
\, \exp [ - (\mu_3R_{ijk}/2)^2/2] \,,
\end{align}
where $r_{ij}=r_i-r_j$ is the relative distance between bosons $i$ and $j$,
and $R_{ijk}=(r_{ij}^2+r_{ik}^2+r_{jk}^2)^{1/2}$.
The strength $V_2^0$ is tuned to unitarity, and $V_3^0$ is tuned to reproduce a 
weakly-bound three-particle state with a binding energy $-E_3$ and an associated
radius ${\bar R}_3 \equiv (-2  m E_3 / \hbar^2)^{-1/2}$.
The introduction of both two- and three-body range parameters $\mu_{2,3}$ allows us
to produce arbitrarily weakly-bound trimers for a given set of interaction 
ranges, which is essential to extract universal physics in the deeply-bound
many-body system.

Specific details of the interaction are not relevant as long as they are
very short-ranged and the ground state can be tuned to a shallow trimer.
In any physical system, the geometric tower
of 
Efimov states at unitarity 
is truncated from below due to 
the range of the interaction.
The 
binding energy of
the would-be next deeper trimer 
is $\simeq (22.7)^2$ larger than that of the calculated ground-state trimer, 
hence the shape of our potentials should produce small
effects for $\mu_{2,3} {\bar R_3} \gg 23$ 
\cite{Bedaque:1998kg,Bedaque:1998km}.
Corrections due to the physical interaction range can be included 
through a two-body potential with two derivatives \cite{vanKolck:1998bw}.

We use Variational and Diffusion Monte Carlo (VMC, DMC) methods
for the solution of the Schr\"odinger equation. 
The trial-state wave functions are of the form
\begin{equation}
\Psi_T \ = \ \prod_i f^{(1)}(r_i)
\prod_{i<j} f^{(2)}(r_{ij}) 
\prod_{i<j<k} f^{(3)}(R_{ijk}) \,,
\end{equation}
with 
$f^{(1)}(r) = \exp(-\alpha r^2)$, 
$f^{(2)}(r) = K\tanh(\mu_J r)\cosh(\gamma r)/r$,
and $f^{(3)}(R) = \exp[u_0\exp(-R^2/(2 r_0^2))]$.
The parameters $K$ and $\gamma$ are chosen to have $f^{(2)}(d)=1$ and 
$f^{(2)'}(d)=0$ at the ``healing distance'' $d$.
The variational parameters $\alpha$, $\mu_J$, $d$, $u_0$ and $r_0$
are optimized at the VMC level for each system and interaction as described 
in Ref.~\cite{Sorella:2001}, and $\alpha=0$ to simulate uniform matter. 
The VMC wave function is then used as input
for exact DMC calculations, see for example Ref.~\cite{Foulkes:2001}.
The calculated energies are exact subject to statistical and 
time-step errors that can be made arbitrarily small.
Results for the energy are independent of the trial wave function, though
statistical errors may be large for poor choices.
Other properties are extrapolated from the VMC and DMC results,
which we have tested 
using different trial wave functions. The 
extrapolation errors are very small, on the order of a few percent or less,
similar or smaller than the reported statistical errors.

\emph{Clusters:}
Clusters with six or fewer bosons have been studied extensively
in the literature with an emphasis on Efimov physics 
\cite{Efimov:1970zz,Hammer:2006ct,Stecher:2009,Deltuva:2010xd,Gattobigio:2011ey,
vonStecher:2011zz}, for a review see Ref.~\cite{nai17}.
Slightly larger clusters with similar interactions have also been considered 
previously~\cite{vonStecher:2010,Nicholson:2012zp,Kievsky:2014dua,Blume:2015}.  
Universal behavior was found for small clusters up to $N \leq 15$. 
Non-universal behavior beyond this point was attributed to 
finite-range effects. 
For sufficiently small range,
it is expected that clusters will be universal and have a 
binding energy per particle
\begin{equation}
\frac{E_N}{N}  = \xi_B (N) \frac{E_3}{3},
\end{equation}
where $\xi_B (N)$ is a universal function of $N$.

\begin{figure}[tb]
\includegraphics[width=1.0\columnwidth]{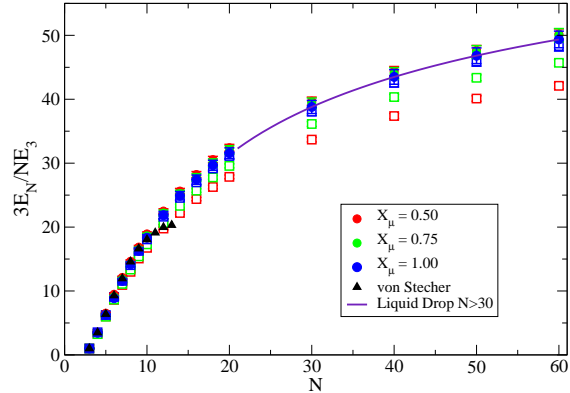}
\caption{Energy per particle of $N$-boson clusters scaled to 
the trimer energy per particle.
Filled symbols are more loosely bound 
($\mu_2 {\bar R}_3  = 65$) and exhibit universal behavior
(the results are also available in~\cite{supp});
open symbols have larger two-body interaction range ($\mu_2 {\bar R}_3  = 46$).
Different colors indicate the ratio of two- to three-body interactions 
ranges, $X_\mu \equiv \mu_3 / \mu_2 =$ 0.5 (red), 0.75 (green) and 1.0 (blue).
Results from Ref.~\cite{vonStecher:2010} are indicated as (black) triangles.
The solid (blue) line corresponds to a liquid-drop fit.}
\label{fig:clusterbinding}
\end{figure}

In Fig. \ref{fig:clusterbinding} we show results for  
clusters of up to 60 bosons for
Hamiltonians with $\mu_2 {\bar R}_3  = 46$ and $65$, and
compare to those of Ref.~\cite{vonStecher:2010} for $N \leq$ 15.
These yield a trimer rms radius 
$\langle r_3^2 \rangle ^{1/2}\approx 0.61 \, {\bar R}_3$ 
for our finite-range Hamiltonians. We consider three-body interactions
with different ratios of two- to three-body interactions ranges, 
$X_\mu \equiv \mu_3 / \mu_2 =$ 0.5, 0.75 and 1.0.
Finite-range interactions will show non-universal effects when
the range of two- or three-particle interactions becomes significant
compared to the average interparticle distance.  This can be seen in
the results of Refs.~\cite{vonStecher:2010,Blume:2015} 
around $N=15$,
and also in our results corresponding to the more bound trimers 
(open symbols with $ \mu_2 {\bar R}_3 = 46$ in 
Fig.~\ref{fig:clusterbinding}) for smaller $X_\mu$.
For $\mu_2 {\bar R}_3 = 65$ the three sets of points with 
$X_\mu = 0.5, 0.75, 1.0$  agree within statistical errors.
For $N=4$ our result ($3E_4/(4E_3)=3.5(1)$ for $\mu_2 {\bar R}_3 = 65$
and $X_\mu = 1.0$) also agrees very well with the precise calculation of 
Ref. \cite{Deltuva:2010xd} ($3E_4/(4E_3)=3.46$), 
suggesting that Efimov-related few-body physics
is properly captured by our potential.

Studies of unitary bosons commonly employ a zero-range
two-body interaction with three-body 
hard-core interaction of radius $R_0$.  
That interaction has a fixed value of 
${\bar R}_3/ R_0 \approx 15.3$ \cite{Comparin:2016thesis},
which can be compared to our 
$\mu_2 {\bar R}_3 = 65 $ and $\mu_3 {\bar R}_3 = 32, 49, 65$ for
$X_\mu = 0.5, 0.75, 1.0$.  The zero-range 
two-body plus 
hard-core interaction can reproduce universal physics for small clusters but 
the three-body hard core is not small compared to typical near-neighbor
separations for larger clusters ($N > 15$) or matter, as discussed below.

For small $N$ the binding energy per particle increases 
approximately linearly with $N$, 
and by $N > 7$ it is an order of magnitude larger than the trimer's.  
Since we have tuned the trimer energy to be very small 
we can find universal behavior up to $N=60$ clusters, 
as shown by the solid points ($\mu_2 {\bar R}_3 = 65$)
in Fig.~\ref{fig:clusterbinding}.
For a 60-particle cluster the 
binding per particle is approximately 50 times that of the trimer.
Naive dimensional arguments would suggest that the repulsive three-body
interaction will become more important for large $N$, resulting
in saturation to a constant binding energy per particle similar 
to what is observed
in atomic nuclei. The energies per particle for large clusters
are beginning to saturate to a constant value as shown in 
Fig.~\ref{fig:clusterbinding}. 
Similar behavior has been seen in finite-temperature simulations
in a trap \cite{Piatecki:2014,Comparin:2016}. 

\begin{figure}[tb]
\includegraphics[width=1.0\columnwidth]{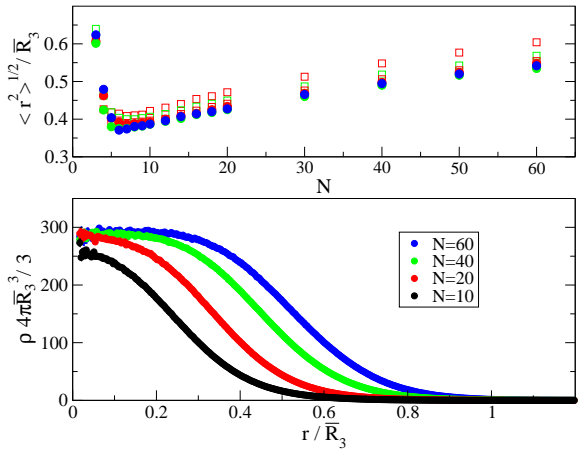}
\caption{Rms radii of $N$-boson clusters (upper panel) and
radial one-body density of various clusters (lower panel),
in units given by the three-boson distance scale ${\bar R}_3$.
Symbols in the upper panel are the same as for Fig. \ref{fig:clusterbinding}.
In the lower panel, 
the curves with $N$= 10 (black), 20 (red), 40 (green) and 60 (blue) bosons 
are for $\mu_2 {\bar R}_3 = 65$ and $X_\mu = 1$.}
\label{fig:clusterradii}
\end{figure}

We have also calculated the single-particle densities and 
radii of the $N$-particle clusters.
Radii
are also expected to scale with a universal ratio  of the trimer rms radius:
$\langle r^2_N \rangle^{1/2} = \beta (N) \langle r^2_3 \rangle^{1/2}$.
Results are shown in Fig.~\ref{fig:clusterradii}.
The upper panel shows that the cluster radius
reaches a minimum around $N$=5-7, and then increases as saturation sets in.
For larger clusters one would expect the radius to increase as $N^{1/3}$
for a system saturating to an equilibrium density.  The lower panel
shows single-particle densities for different particle numbers $N$
and demonstrates the saturation of the single-particle 
density near the center of the clusters at a value 
independent of cluster size.

\begin{figure}[tb]
\includegraphics[width=1.0\columnwidth]{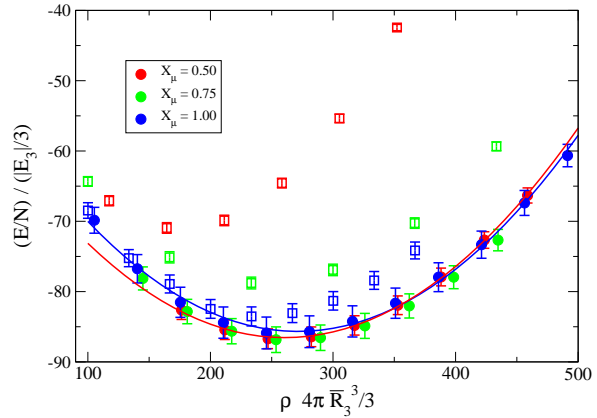}
\caption{Zero-temperature equation of state versus density for the unitary 
Bose fluid. Symbols as in Fig. \ref{fig:clusterbinding}.
The two curves show quadratic fits around saturation density for  
$X_\mu \equiv \mu_3 / \mu_2 =$ 0.5 (red) and 1.0 (blue).}
\label{fig:bosematter}
\end{figure}

\emph{Matter:}
We have also computed the properties of the bulk Bose liquid at 
unitarity for these same interactions using periodic boundary conditions. 
We expected very small finite-size 
effects, and confirmed this by comparing results for 20, 40 and 60 particles.
Results for different 
$N$ at the same density are equivalent
within statistical errors.
We find a universal equation of state (EOS) 
with an equilibrium ground-state energy per particle 
of $87 \pm 5$ times that of the trimer, and a saturation density
of $\rho_0 4 \pi {\bar R}_3^3 /3 = 275 \pm 20$.
The results are summarized in Fig. \ref{fig:bosematter}.
Near saturation density they are well described by 
\begin{equation}
\frac{3E_N (\rho)}{N|E_3|} |_{N\rightarrow\infty}  = \xi_B (N\rightarrow \infty)
\left[-1+ \kappa \left(\frac{\rho-\rho_0}{\rho_0}\right)^2\right],
\label{Evsrho}
\end{equation}
with the dimensionless compressibility $\kappa= 0.42 (5)$.
The curves in Fig. \ref{fig:bosematter} are fits to the EOS calculations
with two different $X_\mu$.

The calculations of the liquid are consistent with those obtained by
extrapolating the cluster results.
A liquid-drop extrapolation of the cluster binding energies,
$E_N/N  = E_B ({N\rightarrow \infty}) ( 1 + \eta N^{-1/3} + 
\ldots )$,
is consistent with the energies found for the bulk. 
Fitting results for $N > 30$, we find that
the universal energy parameter $ \xi_B (N \rightarrow \infty) = 90 \pm 10$. 
The surface energy scaled by the volume energy $E_B ({N\rightarrow \infty})$
is $\eta = - 1.7 \pm 0.3$, but has relatively
large statistical errors.
Similarly, the single-particle density 
near the  center of the drops shown in Fig. \ref{fig:clusterradii}
is consistent with the equilibrium density of matter.
Despite the growth in energy with $N$,
the liquid can be considered universal: the interparticle
separation at equilibrium, 
$[3/(4\pi \rho_0)]^{1/3}\simeq {\bar R}_3/6.4$, is almost
four times larger than the distance scale set by the 
next deeper Efimov trimer in the universal system without cutoffs.
It is also 5-10 times larger than the two- and three-body
interaction ranges, in contrast to 
$\sim 2$  for a zero-range two-body plus three-body 
hard-core interaction at the same density.

It is interesting to compare these results to liquid $^4$He, which 
has a large two-body scattering length and, for small $N$,
weakly-bound clusters 
that can be described by short-range
interactions.  Per particle, 
the binding energy of liquid $^4$He is $-7.14$ K~\cite{Roach:1970}, which is
about 180 times that 
of the $^4$He trimer, $-0.0391$ K~\cite{Pandharipande:1983}.
The scaled surface energy is $\approx - 2.7$~\cite{Pandharipande:1983} and the 
dimensionless compressibility is
$\approx 1.9$~\cite{Roach:1970}.
For small $N$ the helium clusters are universal \cite{Bazak:2016wxm}, 
but for large $N$ the
interaction range is comparable to the interparticle separation and hence
not universal. Nevertheless, the ratio of binding energies 
$\xi_B (N\to \infty)$ 
and the scaled surface energy $\eta$ are 
within a factor of 2 of unitary bosons.

\begin{figure}[tb]
\includegraphics[width=1.10\columnwidth]{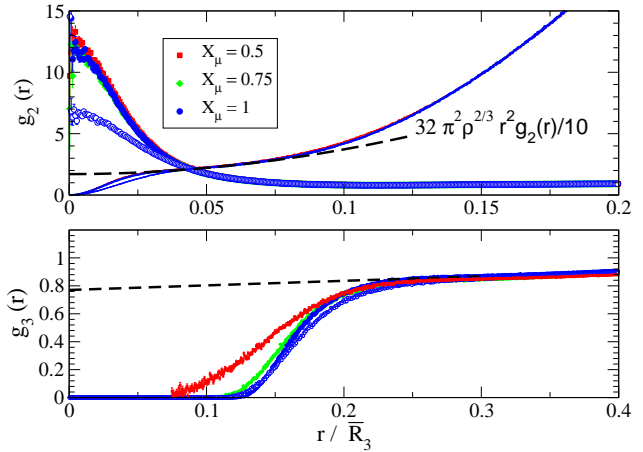}
\caption{Two- (upper panel) and three- (lower panel) body distributions in the
unitary Bose fluid at saturation density, both normalized to unity 
at large separations.
Symbols indicate simulations with different 
two- and three-body ranges, as in previous figures.  
For the two-body distributions in the upper panel, multiplying
by $r^2$ allows an extrapolation (dashed line) to $r=0$ to obtain the
contact.  In the lower panel, the three-body contact (dashed line) can
also be extrapolated from the 
universal regime (dashed line) to $r=0$.
}
\label{fig:bosedist}
\end{figure}

We have also examined the two- and three-body contact parameters $C_{2,3}$
for the unitary Bose fluid at equilibrium density.  
These contact parameters impact
various properties of the system, and relate the short-distance behavior to the
high-momentum tail of the momentum distribution, see for example 
Refs. \cite{Smith:2013,Braaten:2011ur,Werner:2010er,Werner:2012,Gandolfi:2011}.
The two- and three-body
distribution functions are
shown in Fig. \ref{fig:bosedist}, normalized 
to one at large distances (differing by a factor of $ \rho N!$ from 
the $g_N$ defined in Refs. \cite{Werner:2010er,Werner:2012}).

In the universal regime outside the range of the
interaction, the two-body distribution $g_2 (r)$, with $r\equiv r_{ij}$,
is expected to be proportional to $1/r^2$.
The upper lines
in the top panel show $ 32 \pi^2 \rho^{2/3} r^2 g_2 (r) / 10 $ for the different
simulations, and the dashed line is a quadratic fit to results in the universal
regime that can be extrapolated to $r=0$ to give the dimensionless 
two-body contact $\alpha_2$, with $C_2 = N \alpha_2 \rho^{4/3}$ \cite{Smith:2013}.
From the extrapolation of $r^2 g_2$  we find
$\alpha_2 = 17(3)$. More accurate results may be achievable through
simulations at different scattering lengths with fixed $E_3$.
This result is larger but qualitatively comparable to those obtained in more 
approximate approaches \cite{vanHeugten:2013} or those obtained with
zero-range two-body plus 
hard-core three-body interactions \cite{Rossi:2014,Sykes:2014}, 
and quite similar to those extracted through rapid experimental quenches 
\cite{Braaten:2011ur, Smith:2013}.

Similarly, in the bottom figure the dashed line is a fit to $g_3 (r)$,
with $r\equiv R_{ijk}$.
In the universal regime, extrapolating to $r=0$ gives the three-body contact.
It is more accurate to extract the dimensionless three-body contact 
$\beta_3$, with $C_3= N\beta_3 \rho^{2/3} $ \cite{Smith:2013},
from the
derivative of the energy with respect to ${\bar R}_3$ at constant scattering
length. Using the equilibrium properties calculated in Fig. \ref{fig:bosematter}
we obtain $\beta_3 = 0.9(1)$.
The density dependence of $\beta_3$ around equilibrium can be extracted
from Eq. \eqref{Evsrho}.
Further simulations could yield the density dependence of $\alpha_2$,
and also the asymptotic behavior of the momentum distribution $g_3$.

With these strong correlations  and the large binding and small radii 
relative to the trimer, one might expect that the condensate 
fraction may be reduced in the bulk.  In fact we find quite a large
condensate fraction at equilibrium density, with 
$n (k=0)  = 0.93 (1)$,
compared to a value of unity for a weakly-interacting Bose fluid.
One can also compare to liquid $^4$He which has a condensate fraction
of  $0.0725(75)$ at equilibrium density \cite{gly11}.

The large condensate fraction implies that it is reasonable to access
equilibrium properties of the universal Bose fluid as a function of density
through experiments with rapid quenching of a weakly-interacting 
gas \cite{Makotyn:2014,Fletcher:2017}.  
The universal properties of the
unitary Bose fluid are difficult to measure using standard techniques
because of losses to deeply-bound three-body states that occur in
cold atoms but are absent in our simulations. These loss
mechanisms can lead to a trap lifetime smaller than that needed to reach
full equilibrium, and presently
available studies investigate this dynamics of
the rapid quenching of the free-to-unitary transition.
Our results indicate that a rapid quench from a weakly-interacting Bose fluid 
at the appropriate density may enable one to obtain
the equilibrium properties. The relatively large
overlap of the two states should lead to a rapid ejection of particles
through high-energy two- and three-body processes, leading to a rapid
cooling of the system.  Quantifying this energy loss could lead to an
experimental verification of the universal properties of the unitary
Bose fluid in thermal equilibrium.

\emph{Summary:}
We have  demonstrated the
universal nature of bosons at unitarity using short-range interactions
tuned to unitarity in the two-body system and weak binding (Efimov) trimers
in the three-body system.
We have determined many of the universal properties of the
unitary Bose fluid, including
the energies and radii of clusters of up to 60 bosons and
calculated the universal saturation and contacts of the unitary Bose fluid.
We find a ground-state energy per particle of approximately 
90 times that of the trimer at an associated high density.  
We find a large two-body contact parameter, yet 
the condensate fraction in the bulk is greater than 90\%. We also calculate
the three-body contact parameter for the first time.
Further experimental and theoretical studies of the 
unitary Bose fluid will be very intriguing.  Many new properties can be
studied, including those described above, collective effects and the
static and dynamic response of the system.

\emph{Acknowledgments:}
We thank Daekyoung Kang for many valuable discussions.
The work of J.C. and S.G. was supported by the NUCLEI SciDAC program,
and by the U.S. DOE under contract DE-AC52-06NA25396.
The work of U.vK. was supported in part
by the U.S. Department of Energy, Office of Science,
Office of Nuclear Physics, under award number DE-FG02-04ER41338,
and 
by the European Union Research and Innovation program Horizon 2020
under grant agreement no. 654002. 
S.V. thanks the hospitality and financial support from LANL and the
facilities offered by CENAPAD-SP.
Computational resources have been provided by Los Alamos Open
Supercomputing.
We also used resources provided by NERSC, which is supported by the US
DOE under Contract DE-AC02-05CH11231.

\bibliographystyle{apsrev}

\end{document}